\documentclass[]{spie}
\usepackage{graphicx}
\DeclareGraphicsExtensions{.pdf,.png,.jpg}
\usepackage[caption=false%,labelformat=parens,subrefformat=subparens,listofformat=subparens
]{subfig}
\usepackage{lipsum}
\usepackage{wrapfig}
\usepackage{booktabs}
\usepackage{lmodern}
\usepackage[cmex10]{amsmath}
\usepackage{tabularx}
\usepackage{multirow}
\usepackage{booktabs}
\usepackage{color}
\usepackage{changepage}
\usepackage{array}

\newcommand{\beq}{\begin{equation}}
\def\eeq{\end{equation}}
\title{Segmentation-free x-ray energy spectrum estimation for computed tomography}

\author{Wei Zhao\supit{a}, Qiude Zhang\supit{a}, Tianye Niu\supit{b}
\skiplinehalf
\supit{a}Department of Biomedical Engineering, Huazhong University of Science and Technology, Hubei, China 430074;\\
\supit{b}Sir Run Run Shaw Hospital, Zhejiang University School of Medicine; Institute of Translational Medicine, Zhejiang University, Hangzhou, Zhejiang, China 310016.
}

\begin{document}

\maketitle

\begin{abstract}

X-ray energy spectrum plays an essential role in imaging and related tasks. Due to the high photon flux of clinical CT scanners, most of the spectrum estimation methods are indirect and are usually suffered from various limitations. The recently proposed indirect transmission measurement-based method requires at least the segmentation of one material, which is insufficient for CT images of highly noisy and with artifacts. To combat for the bottleneck of spectrum estimation using segmented CT images, in this study, we develop a segmentation-free indirect transmission measurement based energy spectrum estimation method using dual-energy material decomposition.
The general principle of the method is to compare polychromatic forward projection with raw projection to calibrate a set of unknown weights which are used to express the unknown spectrum together with a set of model spectra. After applying dual-energy material decomposition using high- and low-energy raw projection data, polychromatic forward projection is conducted on material-specific images. The unknown weights are then iteratively updated to minimize the difference between the raw projection and estimated projection. Both numerical simulations and experimental head phantom are used to evaluate the proposed method. The results indicate that the method provides accurate estimate of the spectrum and it may be attractive for dose calculations, artifacts correction and other clinical applications.

\end{abstract}

%\keywords{PHS,SYS,XIM}

%%Text from original Abstract
\section{Introduction}

X-ray spectrum plays an very important role in CT imaging, including dose calculation~\cite{demarco2005}, polychromatic image reconstruction~\cite{elbakri2002}, artifacts reduction~\cite{zhao2014,zhao2015sac}, material decomposition~\cite{long2014}, energy-resolved imaging and etc.
In clinical applications, the x-ray flux is usually quite high in order to meet the fast imaging requirement. Thus it is not easy to directly measure the energy spectrum of a CT scanner using an energy-resolved detector, as the detector count rate is usually limited and the pile-up effect is severe. Instead, spectrum calibration often employs indirect methods, including Compton-scattering measurement~\cite{duisterwinkel2015}, Monte Carlo (MC) simulation~\cite{bazalova2007}, empirical or semi-empirical physical models~\cite{tucker1991} and transmission measurements~\cite{sidky2005,duan2011}.

The accuracy of these methods is usually suffered from various limitations. For example, environment conditions (such as low temperature requirement) or hole trapping effect which yields low-energy tailing may affect the spectrum measured using energy-resolved detectors~\cite{koenig2012}. Attenuation and scattering (e.g. Rayleigh and multiple Compton) in the material of the scatterer of the Comton-scattering measurement need to be carefully considered. Transmission measurements based on step or wedge phantom requires dedicated hardware or workflow. Indirect transmission measurements (ITM)~\cite{zhao2015} needs at least the segmentation of one material class. When noise or artifacts are present in the reconstructed image, it causes incorrect material segmentation and yields inferior estimate of the spectrum.

This work aims to develop a segmentation-free indirect transmission measurement-type energy spectrum estimation method using dual-energy material decomposition. Different from ITM~\cite{zhao2015} where polychromatic forward projection is conducted on segmented images, the herein proposed method performs polychromatic forward projection using material-specific images. Thus the method can be applied to estimate spectrum where CT image segmentation is tough. %With this procedure, segmentation issue can be directly avoided.

\section{Methods}
%%-----------------------------------------------------------
\subsection{Workflow of the proposed algorithm}

To avoid determining each energy bin of the X-ray spectrum, we use model spectra to express the spectrum that is to be estimated. The model spectra expression can significantly reduce the degree of freedom of the spectrum estimation problem. In this case, the unknown spectrum $\Omega(E)$ is the weighted summation of a set of model spectra $\Omega_{i}(E)$, i.e.
\begin{equation}\label{equ:spek}
\Omega(E)=\sum_{i=1}^{M}c_{i}\Omega_{i}(E),
\end{equation}
with $M$ the number of the model spectra and $c_i$ the weight on the respective model spectrum. The model spectra can be predetermined using spectrum generators (such as SpekCalc~\cite{poludniowski2009} and Spektr~\cite{siewerdsen2004}) or MC simulation toolkits.

The flowchart of the proposed algorithm is presented in Figure 1. The method starts from acquiring dual-energy raw projection data $p_m$ (i.e., low-energy data $p_L$ and high-energy data $p_H$), based on which, material-specific images are obtained by using either projection-domain or image-domain material decomposition algorithms. The material images are then employed along with the model spectra expression to calculated a set of estimated projection $\hat{p}$. By iteratively updating the unknown weights $c_i$, we can converge to a set of optimal $c_i$ to minimize the quadratic error between the measured raw projection $p_m$ (either $p_L$ or $p_H$) and the estimated projection $\hat{p}$. The unknown spectrum is finally yielded by using Eq (\ref{equ:spek}). The three major components of the approach will be detailed in the following subsections: dual-energy material decomposition, material image-based polychromatic reprojection, and weight estimation.%calibration of unknown weights of the model spectra.

\begin{figure}[t]
    \centering
    \includegraphics[width=0.8\textwidth]{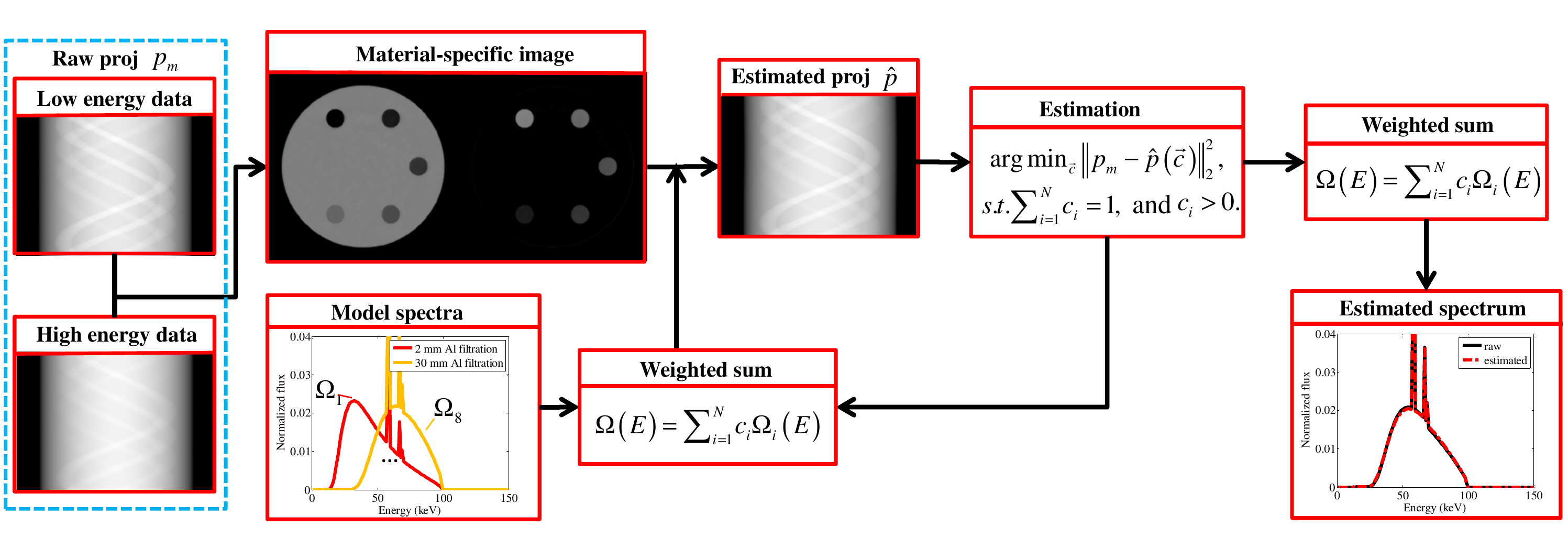}
    \caption{Flowchart of the proposed dual-energy material decomposition-based spectrum estimation method.}
    \label{fig:f1}
\end{figure}

\subsubsection{Dual-energy material decomposition}
Since magnified noise is a general concern for both projection-domain and image-domain dual-energy material decomposition, in this study, to keep the accuracy of the estimated projection $\hat{p}$, we have used an iterative image-domain method to obtain the noise significantly reduced material-specific images~\cite{niu2014}.

\subsubsection{Polychromatic projection on decomposed material images}

In dual-energy material decomposition, the linear attenuation coefficient $\mu(\vec{r},E)$ is modeled with two basis materials via a weighted summation fashion as,
\beq\label{equ:decomposition}
\mu(\vec{r},E)=f_{1}(\vec{r})\psi_{1}(E)+f_{2}(\vec{r})\psi_{2}(E).
\eeq
Here $\psi_{1,2}$ are the known independent energy dependencies which can be mass attenuation coefficients of basis materials and $f_{1,2}(\vec{r})$ are the material-selective images. Based on the above formulation, polychromatic projection of an object is represented as
\beq\label{equ:polyreprojBimg}
\hat{I}=N\int_{0}^{E_{max}}\mathrm{d}E\,\Omega(E) \, \eta(E)\,\mathrm{exp}\left[-A_{1}\psi_{1}(E)-A_{2}\psi_{2}(E)\right],
\eeq%-\sum_{j=1,2}A_{j}\psi_{j}(E)
with $A_{1}=\int_{L}\mathrm{d}\vec{r}\,f_{1}(\vec{r})$ and $A_{2}=\int_{L}\mathrm{d}\vec{r}\,f_{2}(\vec{r})$ the line integral of the material-selective images. Here $L$, $\Omega(E)$ and $E_{max}$ are the propagation path length of each ray, the corresponding polychromatic x-ray spectrum of the ray and the maximum photon energy of the spectrum, respectively. $\eta(E)$ is the energy dependent response of the detector.
Note that $\hat{I}$ is detector pixel dependent and the detector channel index is omitted for convenience. For the absent of the object, the flood field $I_{0}$  can be expressed as follows:
\begin{equation}\label{equ:floodfield}
\hat{I}_{0}=N\int_{0}^{E_{max}}\mathrm{d}E\,\Omega(E) \, \eta(E).
\end{equation}
After applying the logarithmic operation, the projection data can be expressed as:
\begin{eqnarray}\label{equ:esproj}
\begin{split}
\hat{p}(\vec{c})=& \log\left(\frac{\hat{I}_{0}}{\hat{I}}\right) \\
=&\log\left(\frac{\int_{0}^{E_{max}}\mathrm{d}E\,\Omega(E)\,\eta(E)}{\int_{0}^{E_{max}}\mathrm{d}E\,\Omega(E)\,\eta(E)\,\mathrm{exp}\left[-A_{1}\psi_{1}(E)-A_{2}\psi_{2}(E)\right]}\right).
\end{split}
\end{eqnarray}

\subsubsection{Weights estimation}

To estimate the unknown weights for each model spectrum, we minimize the quadratic error between the detector measurement $p_m$ ($p_m$ is either $p_L$ or $p_H$) and the corresponding estimated projection $\hat{p}$ by iteratively updating the weights. This procedure is formulated as the following optimization problem,
\begin{equation}\label{equ:opt-constraint2}
\vec{c}=\underset{\vec{c}} {\mathrm{argmin}}\;\|p_{m}-\hat{p}(\vec{c})\|_{2}^{2},    ~~\mathrm{s.t.}~\sum_{i=1}^{M}c_{i}=1,~\mathrm{and}~c_{i}>0.
\end{equation}
Here the normalization constraint $\sum_{i=1}^{M}c_{i}=1$ and the non-negative constraint which keeps the solution of the problem physically meaning, are introduced. The objective function should be minimal if the spectrum expressed using the model spectra matches the unknown raw spectrum. To solve Eq~(\ref{equ:opt-constraint2}), we use a sequential optimization approach, i.e. minimizing the objective function, followed by normalizing the solution and enforcing non-negative constraint sequentially.

%For each detector pixel measurement, $p_m$, the quadratic error between the measurement and its corresponding estimated projection data, $$\hat{p}$$, should be

%%-----------------------------------------------------------
\subsection{Evaluation}

We first use numerical simulation to evaluate the proposed spectrum estimation method. A water cylinder with six iodine concentrate inserts (range, 0 - 20 mg/mL with 4 mg/mL interval) was simulated in a 2D fan-beam CT geometry. The diameter of the water cylinder is 198 mm and the diameter of the six inserts are 22.5 mm. The low- and high-energy spectra are 100 kVp and 140 kVp, which were generated using the SpekCalc software~\cite{poludniowski2009} with 12 mm Al and 0.4 mm Sn + 12 mm Al filtration, respectively. For the x-ray detection, an energy integrating detector is simulated with 0.388 mm pixel size and 1024 pixels. The x-ray source to the isocenter distance and to the detector distance are 785 mm and 1200 mm, respectively. A set of 720 view angles were scanned in an angular range of $360^0$. Since one difficulty of DECT decomposition is the ill-conditioning, Poisson noise was included in the raw projection to show the robustness of the algorithm. In addition, first order beam hardening correction was performed to improve the accuracy of the material-specific images.

The algorithm was also evaluated using experimental data from an anthrophomorphic head phantom scanned on a cone-beam CT (CBCT) benchtop system. The distance of source to isocenter and source to detector are 1000 mm and 1500 mm, respectively. A total of 655 projections were evenly acquired in 360 degree rotation with $2\times2$ rebinning mode and narrow collimation to avoid scatter radiation. Tube potentials of high and low-energy spectra were 125 kVp and 75 kVp, respectively. Both of the spectra were filtered with a 6 mm aluminium filter.

For both of the numerical simulation and experimental evaluations, low and high energy CT images were reconstructed by using a filtered backprojection (FBP) algorithm with the band-limited Ramp filter (i.e. Ram-Lak filter) whose cut-off frequency is set to the Nyquist frequency. Low-energy data sets are used to estimate low-energy spectra and high-energy spectra estimation is exactly the same.

To quantify the accuracy of the estimated spectrum, we calculate the normalized root mean square error (NRMSE) and the mean energy difference $\Delta E$ between the raw spectrum (ground truth) and the estimated spectrum, i.e.

\begin{equation}\label{equ:error}
NRMSE=\sqrt{\frac{\sum_{e=1}^{N}(\hat{\Omega}(e)-\Omega(e))^{2}}{\sum_{e=1}^{N}\Omega(e)^{2}}}    %\frac{\|\hat{\Omega}-\Omega\|_{2}}{\|\Omega\|_{2}}.
\end{equation}
\begin{equation}\label{equ:meanEnergy}
\Delta E= \sum_{e=1}^{N}E_e\,(\Omega(e)-\hat{\Omega}(e)),
\end{equation}

with $\hat{\Omega}(e)$ the $e$th energy bin of the normalized estimated spectrum and $\Omega(e)$ $e$th energy bin of the normalized true spectrum. $N$ and $E(e)$ are the number of the energy bins and the energy of the $e$th energy bin of the spectrum, respectively.

\section{Results}
\subsection{Numerical simulation}

\begin{figure}%%[ht]
    \centering
    \includegraphics[width=0.5\textwidth]{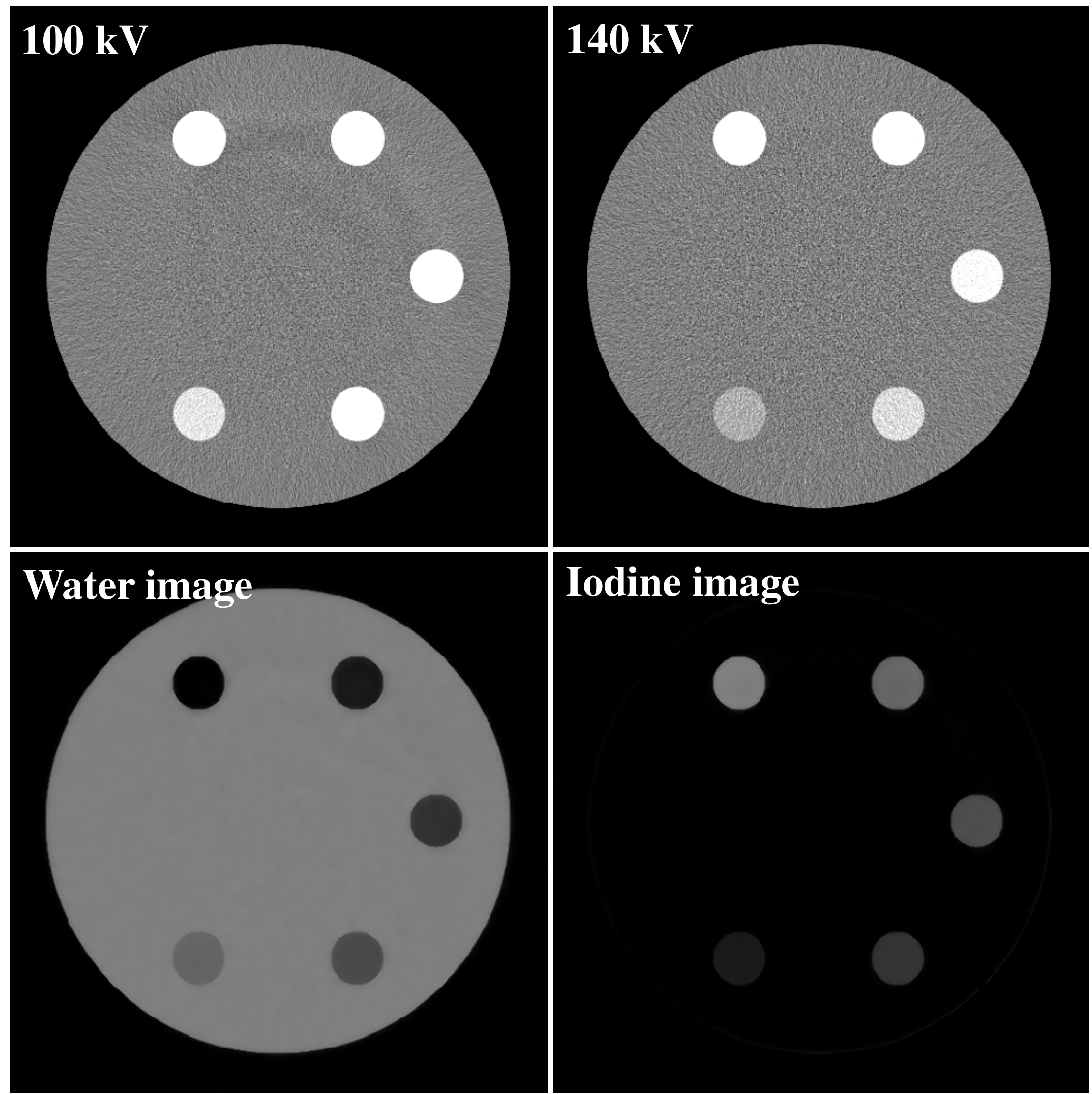}
    \caption{High and low-energy CT images and material-specific images of the numerical iodine concentrate phantom. Display windows: kV CT images, C/W=0 HU/300 HU; material images, C/W=100\%/200\%. }
    \label{fig:f2}
\end{figure}

%\begin{figure}%%[ht]
%    \centering
%    \includegraphics[width=0.5\textwidth]{fig3.pdf}
%    \vspace{-0.5em}
%    \caption{material-specific images generated using iterative image-domain dual energy material decomposition method. }
%    \label{fig:f2}
%\end{figure}

\begin{figure}[t]
    \centering
    \includegraphics[width=0.5\textwidth]{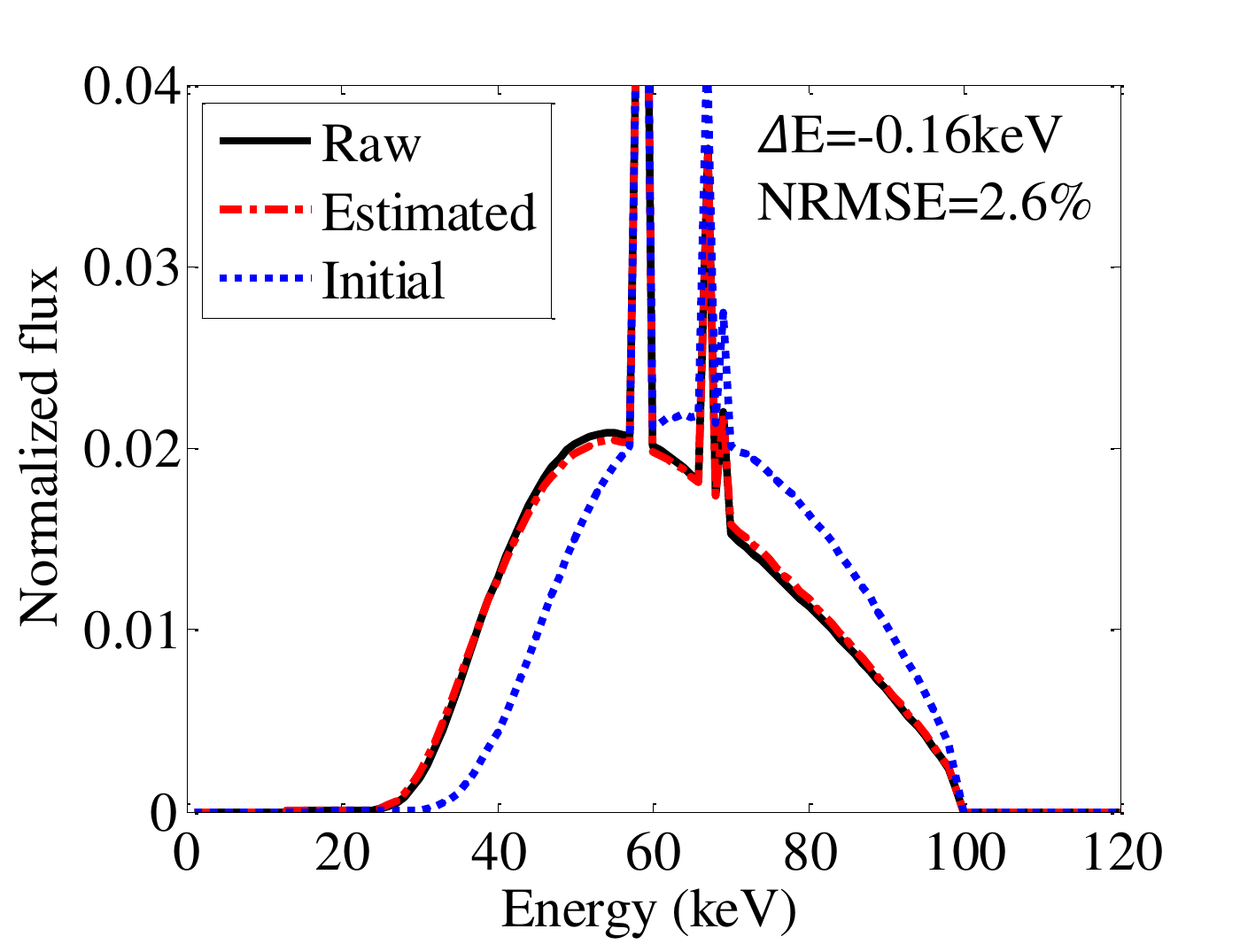}
    \caption{Spectrum estimated using the numerical iodine concentrate phantom. }
    \label{fig:f3}
\end{figure}

Fig.~\ref{fig:f2} shows the results of high and low-energy CT images, and basis material images of the numerical iodine concentrate phantom. As can be seen, the 100 kV image shows much high contrast level for the iodine inserts. Although water correction has been applied, there are some residual high order streaks in the 100 kV images since its spectrum is much softer than the 140 kV spectrum. Superior water and iodine images were obtained by selecting the central of water cylinder and the 20 mg/mL iodine concentrate insert as ROIs to calculate the decomposition matrix. Fig.~\ref{fig:f3} depicts the results of the 100 kV spectrum estimation using the numerical phantom. The initial spectrum is the hardest spectrum of the model spectra. The raw spectrum is the spectrum that was used to generate the 100 kV projection data and it can be regarded as the ground truth. The estimated spectrum matches the raw spectrum quite well and their mean energy difference is 0.16 keV, suggesting the dual-energy material decomposition-based method provides accurate spectrum estimate.

%%------
\subsection{Experiments phantom study}

\begin{figure}
    \centering
    \includegraphics[width=0.5\textwidth]{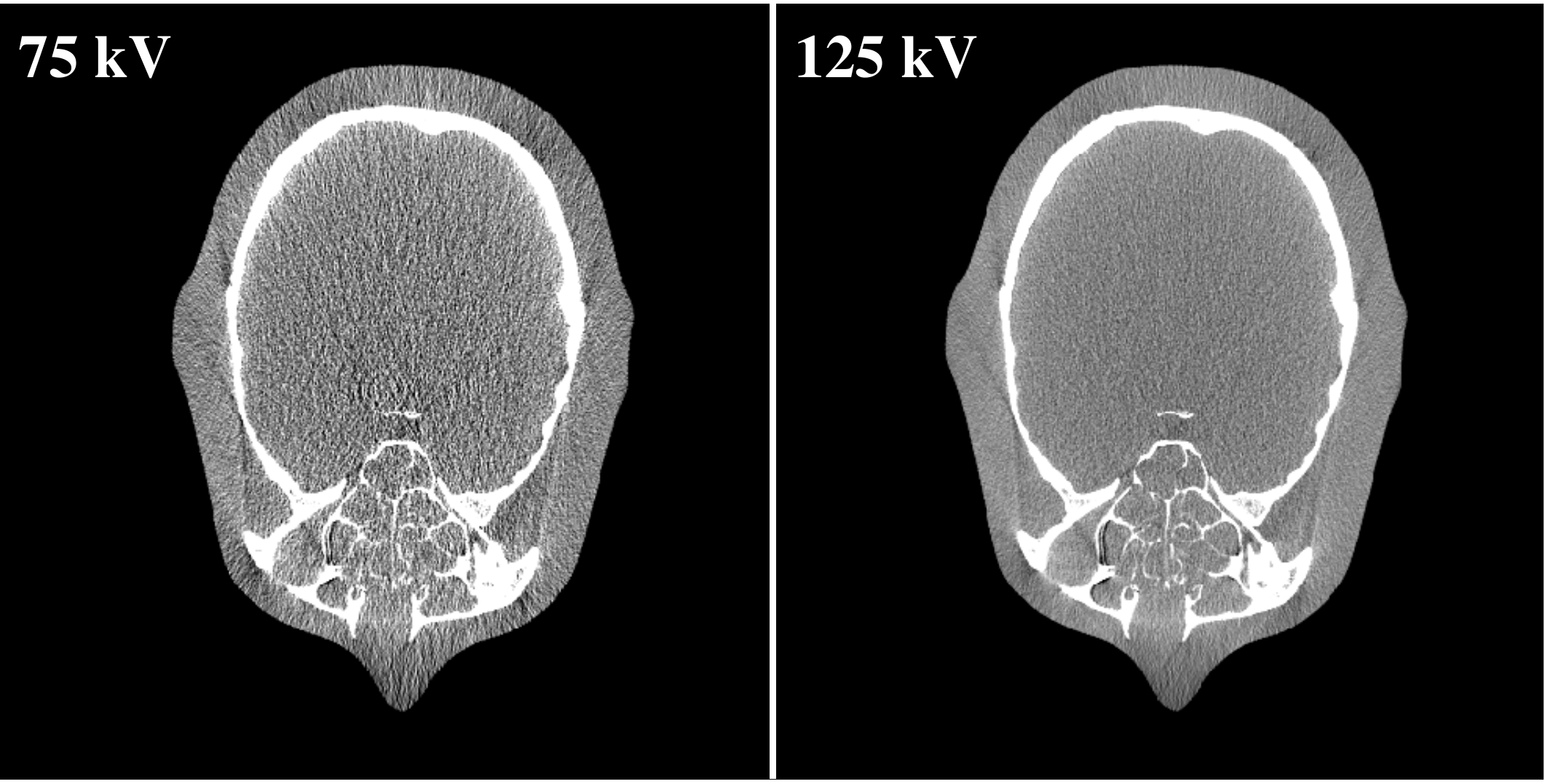}
    \caption{Low- and high-energy CT images of the experimental head phantom. Display window: [-300HU, 300HU].}
    \label{fig:f4}
\end{figure}

\begin{figure}
    \centering
    \includegraphics[width=0.5\textwidth]{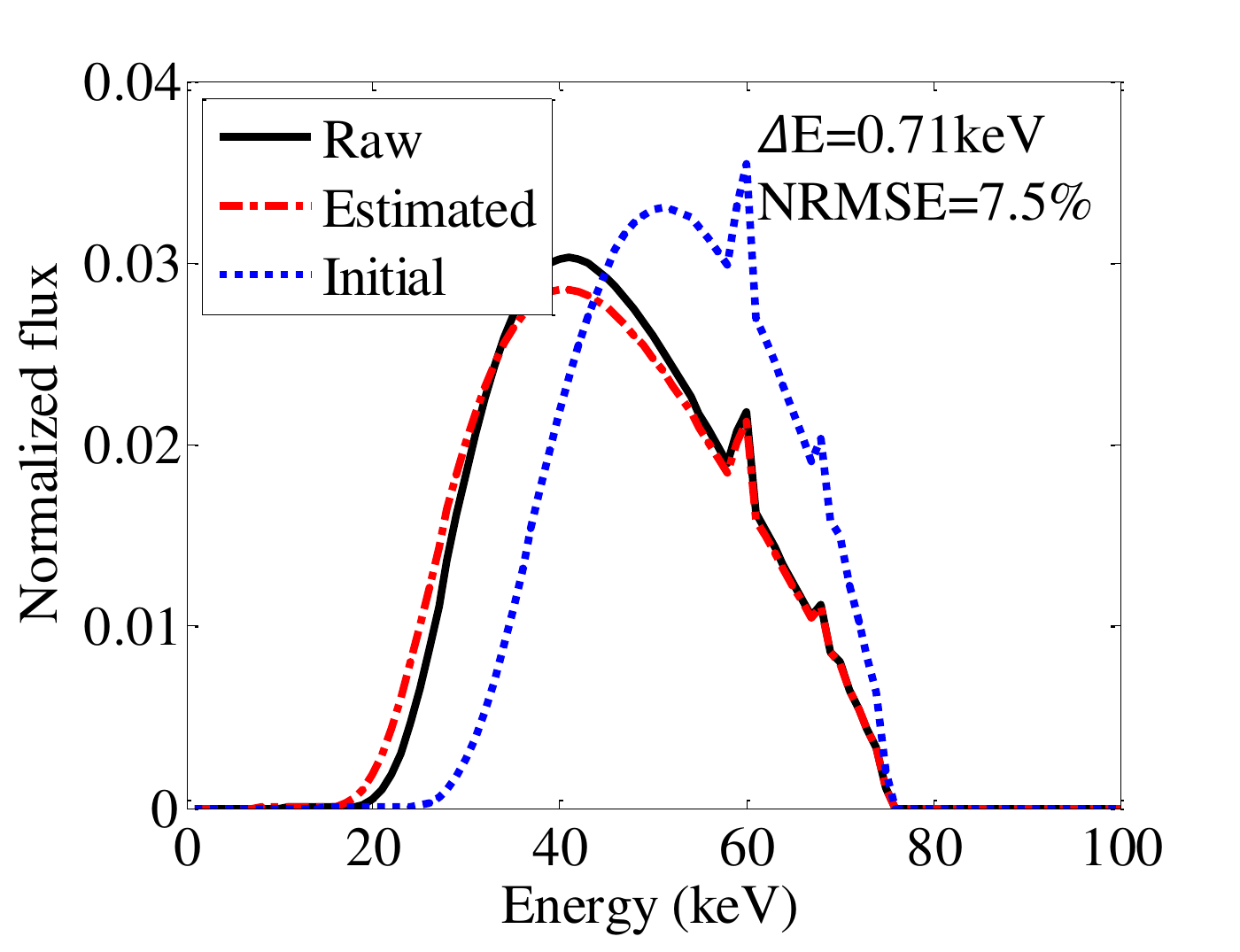}
    \caption{Spectra estimated using the physical head phantom with detector efficiency incorporation. The reference spectrum (Raw) is calculated using SpekCalc software with filtration matched with the experimental setting.}
    \label{fig:f5}
\end{figure}

Fig.~\ref{fig:f4} shows low- and high-energy CT images of the head phantom. For the experimental evaluation, the benchtop CBCT system has used a flat detector with 0.6 mm thickness of CsI. To better estimate the spectrum, energy dependent efficiency has been taken into account. Fig.~\ref{fig:f5} depicts spectrum estimated with the anthrophomorphic head phantom with detector efficiency incorporation. The initial spectrum is the hardest spectrum of the model spectra. As can be seen, the estimated spectrum matches the raw spectrum well. The mean energy difference and NRMSE are 0.71 keV and $7.5\%$, respectively. Note that we do not directly measure the raw spectrum in this case, instead, the well-validated spectrum generator SpekCalc is used to generate the raw spectrum with matched x-ray tube specifications.

%\section{Novelty}
%\begin{itemize}
%	\item{The herein proposed spectrum calibration method does not require dedicated hardware or workflow and it has potential to be conducted on material-specific images yielded by other applications.}
%	\item{Compared to previous indirect transmission measurement-based spectrum estimation method, the proposed method is segmentation-free.}
%	%\item{The proposed method is intrinsically patient-specific.}
%\end{itemize}

\section{Conclusion}
%\vspace{-.5em}
This work presents an x-ray energy spectrum calibration method for CT scanners using dual-energy material decomposition and the indirect transmission measurement framework. The method conducts polychromatic reprojection on material-specific images instead of segmented CT images, with which the segmentation procedure is overcome.  Hence, the herein proposed method does not require dedicated hardware or workflow and it is segmentation-free. The reprojection data is then compared to raw projection data and their difference is minimized by iteratively updating a set of weights, which are used to express the unknown spectrum together with a set of model spectra. The method was evaluated using numerical simulation data and experimental phantom data. The results demonstrate raw spectra can be accurately recovered by incorporating the energy-dependent detector absorption efficiency. Mean energy differences between raw spectra and estimated spectra are 0.16 keV and 0.71 keV for the numerical simulation and experimental phantom data, respectively.

\section*{}

This work has not been submitted to any other conferences or publications.

%\bibliography{spek_estimation}
\bibliographystyle{spiebib}

\end{document}